%% file: main.tex
 \pdfoutput=1
\documentclass[conference]{IEEEtran}
\usepackage[utf8]{inputenc}
\usepackage{cite}
\usepackage{amsmath,amssymb,amsfonts}
\usepackage{algorithmic}
\usepackage{graphicx}
\usepackage{textcomp}
\usepackage{array}
\usepackage{xcolor}
\usepackage{caption}
\usepackage{subfig}
\usepackage{bm}
\usepackage{times}
\usepackage{comment}
\usepackage{fancyhdr,graphicx,amsmath,amssymb}
\usepackage{tikz}
\usepackage{pgfplots}
\pgfplotsset{compat=1.14}
\usetikzlibrary{automata,arrows,positioning,calc}
\usepgfplotslibrary{colorbrewer}

\newcommand{\ilmcom}[1]{\textcolor{teal}{\textsc{\textbf{Comment: }}#1\\}}

\include{pythonlisting}
\def\BibTeX{{\rm B\kern-.05em{\sc i\kern-.025em b}\kern-.08em
    T\kern-.1667em\lower.7ex\hbox{E}\kern-.125emX}}
    
\title{Witness-based Approach for Scaling Distributed Ledgers to Massive IoT Scenarios}

\author{Duc-Lam Nguyen, Israel Leyva-Mayorga, and Petar Popovski \\
\textit{Connectivity Section, Department of Electronic Systems, Aalborg University} \\
Aalborg, Denmark \\
Email: \{ndl, ilm, petarp\}@es.aau.dk
}

\begin{document}
\maketitle

\begin{abstract} 
Distributed Ledger Technologies (DLTs) are playing a major role in building security and trust in Internet of Things (IoT) systems. However, IoT deployments with a large number of devices, such as in environment monitoring applications, generate and send massive amounts of data. This would generate vast number of transactions that must be processed within the distributed ledger. 
In this work, we first demonstrate that the Proof of Work (PoW) blockchain fails to scale in a sizable IoT connectivity infrastructure. To solve this problem, we present a lightweight distributed ledger scheme to integrate PoW blockchain into IoT. In our scheme, we classify transactions into two types: 1) global transactions, which must be processed by global blockchain nodes and 2) local transactions, which can be processed locally by entities called \textit{witnesses}. Performance evaluation demonstrates that our proposed scheme improves the scalability of integrated blockchain and IoT monitoring systems by processing a fraction of the transactions, inversely proportional to the number of witnesses, locally. Hence, reducing the number of global transactions. 
\end{abstract}


\maketitle

\section{Introduction}
Distributed Ledger Technologies (DLTs) provide high levels of security, accountability, tractability, and privacy to the transmitted data~\cite{iotbc}. This is achieved by enabling key functionalities, such as transparency, distributed operation, and immutability~\cite{dac}. The benefits of DLTs are particularly appealing for Internet of Things (IoT) applications, where large amounts of data are generated and the devices can only implement weak security mechanisms~\cite{dorri}.

The trust provided by DLTs is greatly valuable in IoT monitoring applications with a large number of devices. As an example, consider an urban IoT application that monitors the air quality and gas emissions. The data generated by this application is critical, so it must be protected, tractable, immutable, and transparent. Nevertheless, in a traditional monitoring system, the inter-organization sharing the data may be untrusted, complex, unreliable, and non-transparent. Besides, the current IoT-based monitoring systems are centralized, which leads to a single point of failure, where data can be lost or modified~\cite{security}. 


\begin{figure}[t]
    \centering
    \includegraphics[width=0.9\linewidth]{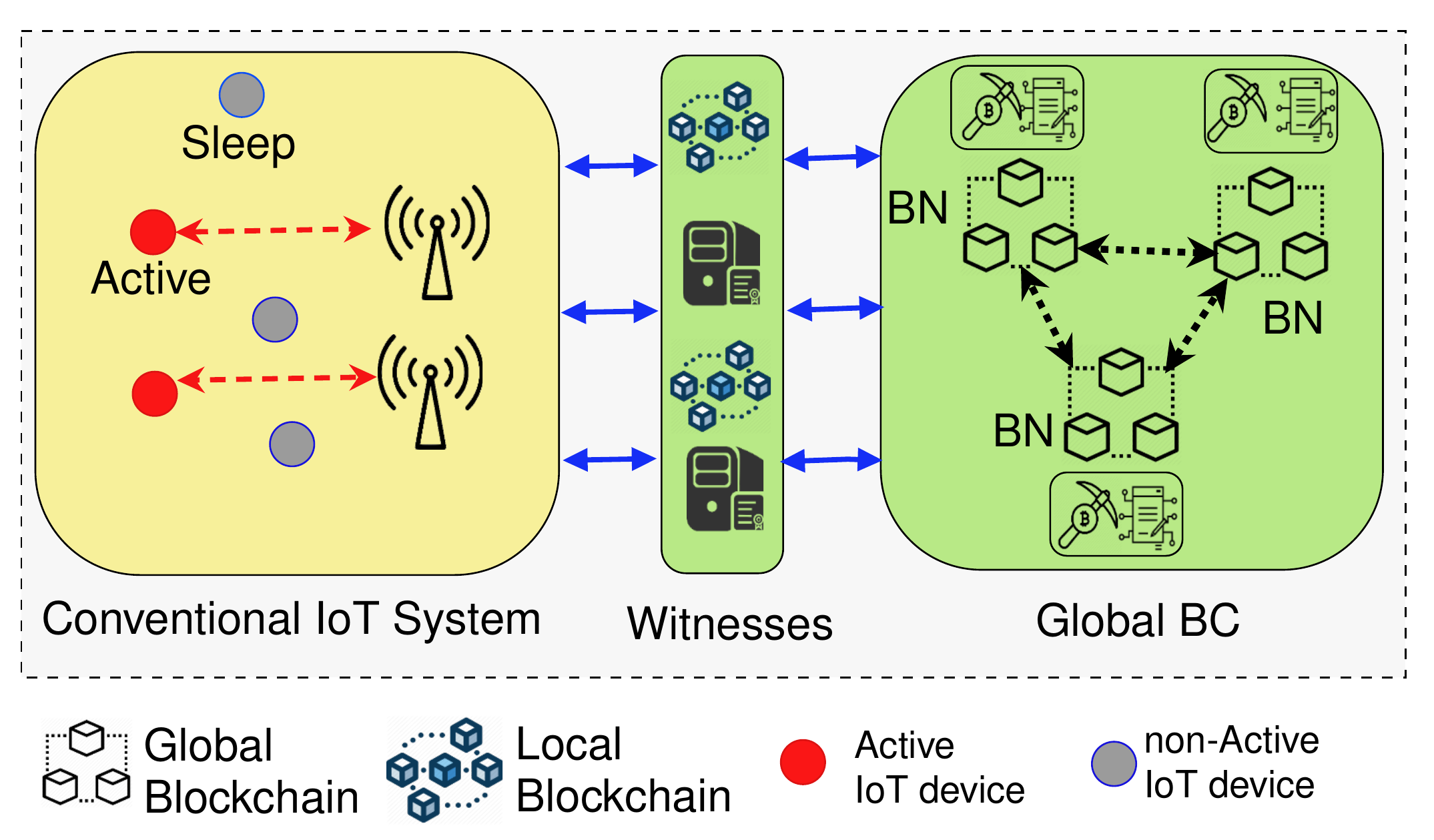}
    \caption{Overview of our Blockchain-enabled IoT system \textit{wiBlock}. The IoT nodes generate and send the transactions to the base stations, which in turn send them to the witnesses. These decide which transactions must be sent to the GB and process the rest.}
    \label{fig:architecture}
\end{figure}

The problems described above may be solved by integrating Blockchain into IoT applications. However, Blockchain architectures were not designed to handle a large number of transactions, which would be generated by naively integrating Blockchain into IoT. Specifically, IoT deployments usually present a star topology, in which the devices communicate directly to the base stations (BS), which then redirects the gathered data to the destination \cite{iotmonitoring} (e.g., from Narrowband IoT (NB-IoT) or LoRa deployments to a cloud server), as shown in the left part of Fig.~\ref{fig:architecture}. In the most Blockchain and IoT integration, this same architecture would be used, and the BS would be in charge of communicating with the Blockchain \cite{lsb}. Thus, every packet generated by the IoT devices would represent a transaction, which can easily overload the Blockchain.

Three main challenges must be overcome to achieve an efficient integration of Blockchain into IoT. First, DLTs use diverse resource-intensive gating functions, for example, \textit{Proof-of-Work (PoW)} and \textit{Proof-of-Stake (PoS)}, while IoT devices are resource-constrained. As a consequence, the processing time of these functions in IoT devices would be restrictive. Second, the widely-used Blockchain arrangement cannot handle the massive transactions generated by IoT devices. For example, Bitcoin network produces $1$~MB blocks, roughly once every 10 minutes, with an average size of transaction around 500 bytes, which give 7 transactions per second (tps). In comparison, Visa system performs $2000$~tps on average, and an average daily peak of $4000$~tps, with a maximum capacity of $56000$~tps. Third, the power saving mechanisms of the IoT devices can cause problems during knowledge dissemination and synchronization. For instance, an update may be severely delayed or even fail to arrive if a device is in sleep mode.

In this paper, we present a witness-based Blockchain system called \textit{wiBlock}, especially designed to integrate Blockchain into resource-constrained IoT applications. It is aimed to solve three of the main problems of traditional IoT monitoring systems, namely trust, scalability, and cost. This is achieved by: 1) enabling the use of DLTs to store IoT data, 2) limiting the number of transactions that must be processed at the 
Global Blockchain (GB), and 3) eliminating the need for complex computations and supporting sleep-awake mechanisms at the IoT devices, respectively.  

The architecture of \textit{wiBlock} is illustrated in Fig.~\ref{fig:architecture}, where the IoT devices interact exclusively with the \textit{witness} system, which then may process the transactions locally or communicate directly with the GB. The transactions that must be processed by the GB are called \textit{global transactions}, whereas the transactions that can be verified locally at the \textit{witness system} are called \textit{local transactions}. In order to see the need for this differentiation, consider a pollution monitoring system, in which a number of sensors in a given local area are associated to the same witness. Then a local transaction can be used to send local sensing data from a device associated with the same witness. For instance, the alarm sensor periodically requests gas sensor which collects the concentration of pollutants e.g., SO2, CO2, NO to detect the abnormal condition in air. In order to see the need for a global transaction, note that sensors may wish to store their sensing data to external storage system e.g, IPFS \cite{ali2017iot} or control a thermostat sensor, which is located in a different area and associated with a different witness to adapt temperature. In this case there is a need to communicate via 
different heterogeneous networks and record the transaction results to the GB via global transactions. Thus, the witness system reduces the number of transactions that need to be processed by the \textit{GB} and the latency of transaction verification. Furthermore, \textit{wiBlock} allows each IoT device to communicate with several witnesses. This avoids having a single point of failure (i.e., bridge) between the IoT device and the GB, which in turn greatly increases the reliability of the IoT application. For example, Blockchain witness models have been found to be beneficial for Cloud Service Level Agreement \cite{witnessInfocom}.

The contributions of this work are as follows: 
\begin{enumerate}
	\item We investigate the possibilities of naively integrating Blockchain directly into resource-constrained IoT systems. We identify some of the major problems that arise in this setup, which illustrate that Blockchain technology is not directly applicable to massive IoT.
	\item We propose a new IoT-friendly distributed ledger system named \textit{wiBlock}. It aims to solve the scalability issues of Blockchain in massive IoT environment by defining two types of transactions: global and local. 
	\item We thoroughly compare the performance \emph{wiBlock} with that of a naive Blockchain and IoT integrated architecture. Our results show that our proposed system enhances the scalability of the GB network. 
\end{enumerate}

The remainder of this paper is organized as follows. In Section~\ref{sec:system_model}, we present the system model, followed by the design of our novel \textit{wiBlock} system in Section~\ref{sec:wiblock}. We present the analysis and performance evaluation of \textit{wiBlock} in Section~\ref{sec:analysis} and Section~\ref{sec:performance}, respectively. Finally, we conclude the paper in Section~\ref{sec:conclusions}.

\section{System model}
\label{sec:system_model}
We consider an IoT application with $k$ devices. These are deployed uniformly at random in a squared area of interest $A\in\mathbb{R}^2$. The IoT devices generate transactions with the data collected from the environment according to a Poisson process with rate $\lambda$. 

In the most simple Blockchain and IoT integrated architecture, the transactions are sent to the BS, which then redirects them to the GB. 
In \emph{wiBlock}, the transactions are sent to the \emph{witness system} instead. This is a set of $v$ witnesses, which have the capacity to verify transactions locally and to communicate with the GB. The time needed for a witness to perform these operations determine its capacity and depend on numerous factors. However, it is out of the scope of this paper to derive their precise values. Transactions are grouped into blocks of size $b$. Therefore, a new block is created when $b$ new transactions are received at a server.

Witnesses may be either physical or logical entities, hence, their organization is flexible. For simplicity, throughout this paper we assume one witness is deployed at each BS and use these terms interchangeably. The BSs are distributed randomly within $A$. We denote the set of IoT devices and witnesses as $\mathcal{D}=\{1, 2, \dotsc, k\}$ and $\mathcal{W}=\{1, 2, \dotsc, v\}$, respectively.

The IoT devices and witnesses communicate through wireless links under a standard path loss model and large-scale (slow) fading. Thus, a transaction is transmitted successfully from IoT device $i$ to a witness $w\in\mathcal{W}$ with probability $p_s(i,w)$. The IoT device $i$ selects the witness $w$ according to a predefined strategy. If the transmission fails, $i$ attempts the transmission to a different witness. This process is repeated until the transaction is confirmed or until a given number of attempts is reached without success. 

We consider a simple shadowing propagation model for the communication between IoT devices and witnesses where, for a given transmission power $P_t$ and carrier frequency $f$, the received power at a distance $d$ is
\begin{equation}
    P_r(d)= 10 \log_{10}\left(\frac{ P_t G_t G_r\,c^2 }{(4\pi f)^2 d^\beta}\right)+N(0,\sigma_\text{dB})~\text{dB}
\end{equation}
where $G_t$ and $G_r$ are the transmitter and receiver antenna gains, respectively,  $c=3\cdot10^8$~m/s is the speed of light, $N(0,\sigma_\text{dB})$ is a zero-mean Gaussian random variable (RV) with standard deviation $\sigma_\text{dB}$~dB, and $\beta$ is the path loss exponent.

From there, the outage probability at a given distance and receiver sensitivity $\gamma$ is
\begin{equation}
    p_\text{out}(d) =1- Q \left ( \frac{1}{\sigma_\text{dB}} 10 \log_{10} \left( \frac{\gamma (4\pi f)^2 d^\beta}{P_t G_t G_r c^2} \right)  \right)
\end{equation}
and $p_s(i,w)=1-p_\text{out}(d(i,w))$. Throughout this paper, we assume that the wireless resources are sufficient to support the communication between the IoT devices and the witness system and do not go into the details of the access protocols. Therefore, collisions caused by simultaneous transmissions from the IoT devices to a witness $w$ can be avoided or resolved if the links toward $w$ are not in outage. Finally, no errors occur in the communication between the witness system and the GB.

\section{\textit{WiBlock} Design}
\label{sec:wiblock}
This section presents the detailed description of \textit{wiBlock} architectural elements and operation. 

\subsection{Witness-based Blockchain System}

 As illustrated in Fig.~\ref{fig:architecture}, the witness-based Blockchain System consists of three main components: the GB, the witness system, and the physical IoT devices. The first action performed by the IoT devices after deployment is authentication. For this, each device $i\in \mathcal{D}$ performs a key exchange procedure with a witness $w\in\mathcal{W}$ to gain the necessary permissions and build secure channels to perform transactions. After authentication, the tuple $(i,w)$ is added by $w$ to the shared registry of the witness system $\mathcal{R}$. 
For this, $w$ shares the authentication information of $i$ (i.e., credentials) with the rest of the witnesses. By keeping a shared registry, device $i$ can communicate with any witness, even though is registered with $w$. After authentication, IoT devices collect the data and sign it by using a \texttt{SecretKey} $s_\text{key}(i)$ that is unique for each $i$ as \textit{\texttt{Sign}(data, $s_\text{key}(i)$, timestamp $\tau$)}. Next, the transaction is created and transmitted to a \textit{witness} $w$. Note that this latter witness may be different to the one which $i$ is registered with. Local transactions, denoted as $L_l$, are exchanged exclusively between $w$ and all the IoT devices registered with it $\{i\in\mathcal{D}: (i,w)\in\mathcal{R}\}$, for which the managers implement a consensus procedure. On the other hand, Global transactions, denoted as $L_G$, must be sent from a \textit{witness} $w$ to the GB when $(i,w)\notin\mathcal{R}$. These two types of transactions are further described in the following.

\begin{figure}[t]
    \centering
    \includegraphics[width=0.7\linewidth]{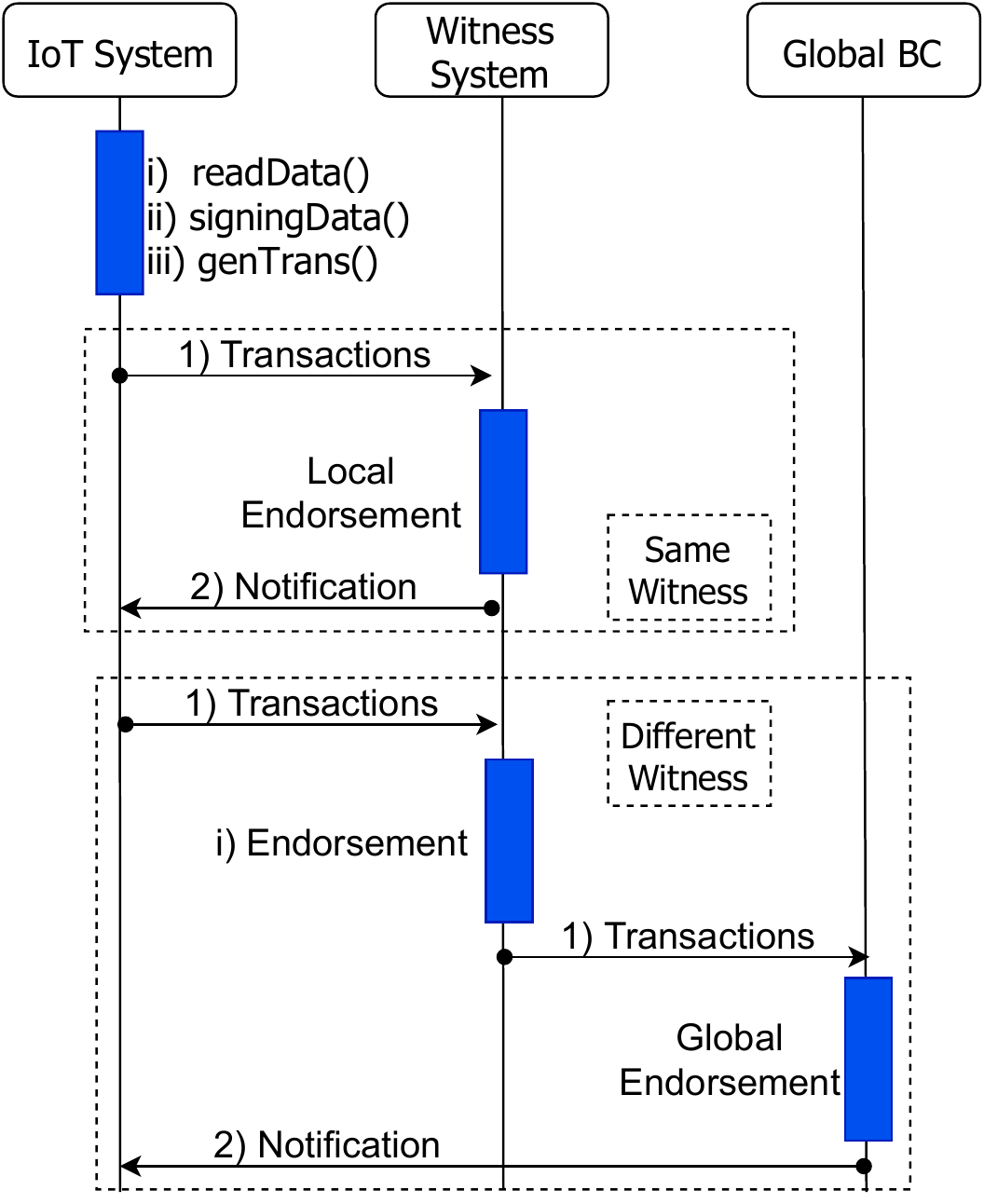}
    \caption{Transaction flow in \textit{wiBlock}, from generation to confirmation.}
    \label{fig:logical}
\end{figure}

\subsubsection{Local Transactions} These transactions are transmitted from IoT device $i$ to the witness $w$, whose key management component confirms that $(i,w)\in\mathcal{R}$. Then, this same component checks whether the \texttt{PublicKey} $p_\text{key}(i)$ of $i$ has been associated with any block in the \textit{local ledger}. If $p_\text{key}(i)$ has not been associated with any block, the \textit{witness} $w$ generates a new block for the given $i$. Then, $w$ arranges the transactions in order, updates the \textit{local ledger}, and a notification feedback message is transmitted to the devices.

\subsubsection{Global Transactions} These transactions are transmitted from IoT device $i$ to the witness $w$, whose key management component confirms that $(i,w)\notin\mathcal{R}$. Then, this same component will clarify which \textit{witness}  $i$ is registered with. If $\exists w'\in \mathcal{W}$ s.t. $(i,w')\in \mathcal{R}$, the transaction is forwarded to the GB. In case the GB has a block associated with given device $i$, the transaction will be validated based on the corresponding signature \textit{\texttt{Sign}(data, $s_\text{key}(i)$, timestamp $\tau$)} and, if the signature is valid, the transaction is appended to the block and transmitted back to the \textit{witness} $w'$. Note that this type of transactions will be frequently generated when the IoT devices are mobile. For example, cargo, supply chain, and car subsystem monitoring.


\subsection{Witness Selection}

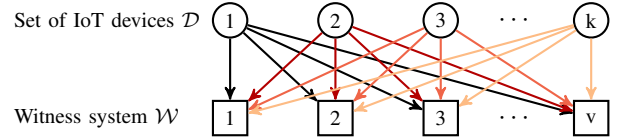
\begin{figure}[t]
	\centering
	\input{figures/witness_selection.tikz}
	\caption{In \emph{wiBlock}, each IoT device has a list of eligible witnesses. Transactions generated by the IoT devices are sent to a witness in this list, according to the witness selection strategy.}
	\label{fig:witness_selection}
\end{figure}

Numerous witness selection strategies can be implemented at the IoT devices and each one may offer different benefits. However, the focus of the present work is to evaluate the benefits of the witness-based architecture, rather than to identify an optimal witness selection strategy. Therefore, we consider the following a heuristic witness selection strategies and evaluate the performance of the witness system. As illustrated in Fig.~\ref{fig:witness_selection}, IoT devices select one of the $v$ available witnesses with probability $1/v$ and transmit the transaction. Then, if the link between IoT device $i$ and witness $w$ is not in outage, the transaction is confirmed. Otherwise, $i$ selects a new witness uniformly at random from $\mathcal{W}\setminus w$ and transmits the transaction. This process is repeated until the transaction is confirmed or until a given number of attempts $l\leq v$ is reached without success. This is the simplest strategy and assumes the IoT devices have no information about the state of the wireless channel toward each witness separately. 

\section{Analysis}
\label{sec:analysis}
\subsection{Queuing model of the witness system}
\begin{figure}[t!]
	\centering
	\includegraphics[width=0.9\linewidth]{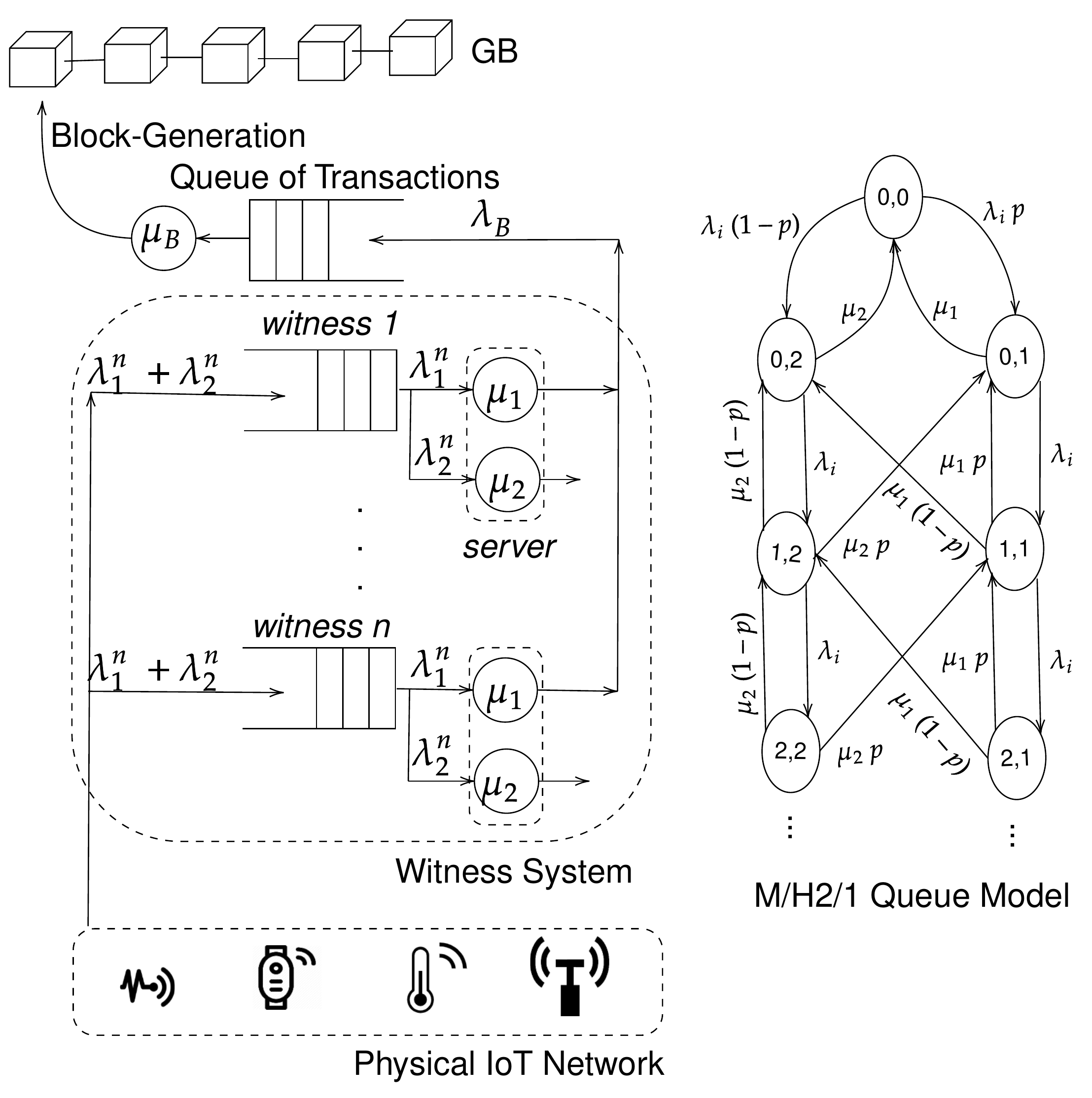}
	\caption{ Witness-based Blockchain queuing model described in Section~\ref{sec:analysis}.}
	\label{fig:witness}
\end{figure}

We consider a queuing model for witness-based Blockchain network as described in Fig.~\ref{fig:witness}. The witnesses and Blockchain are modelled as queuing nodes to capture the number of transactions that must be i) processed locally by witnesses and ii) forwarded to the GB to be processed. We assume that transactions are generated by the IoT devices following a Poisson process. Hence, we denote $\lambda(i)$ as the transaction generation rate at IoT device $i$. 

Let $p(i,w)$ be the probability that $i$ chooses witness $w$ and $p_s(i,w)$ be the probability that the link between $i$ and $w$ is not in outage. Building on this, the average transaction arrival rate at the witness $w$ is 
\begin{equation}
    \lambda_w=\sum_{i=1}^k p(i,w)p_s(i,w)\lambda(i).
\end{equation}
Hence, the transaction arrival rate of different witnesses depends on the density and location of the deployed IoT devices and witnesses, but also on the witness selection criteria. 

The probability $p(i,w)$ depends on the witness selection strategy. For the strategy 1, random selection, let $A(w,u)$ be the matrix of permutations of $u$ elements taken from $\{1-p_s(i,w')\}_{w'\in\mathcal{W}\setminus w}$ with $_{(v-1)}P_u$ rows and $u$ columns. The element in row $x$ and column $y\leq u$ of $A(w,u)$ is denoted $a_{xy}(w,u)$. From there, we can calculate $p(i,w)$ as:
\begin{equation}
    p(i,w)=\frac{1}{v} + \frac{1}{v!}\sum_{u=1}^{l-1}(v-u-1)! \sum_{x=1}^{_{(v-1)}P_u}\prod_{y=1}^{u}a_{xy}(w,u)
\end{equation}

As mentioned above, generated transactions are either \textit{global} $L_G$ or \textit{local} $L_l$. We define $p$ as the probability that a transaction sent to a witness is \textit{Global}. Hence, $1-p$ is the probability that a transaction is \textit{local}. Please observe that the value of $p$ only depends on the number of witnesses $v$ and is $p:=\Pr\left[(i,w)\notin \mathcal{R}\right]=(v-1)/v$. 

 The transaction processing time is assumed to follow an exponential distribution with service rates $\mu_1$ and $\mu_2$ for \textit{global} and \textit{local} transactions, respectively, and transactions are served according to a first-come first-served (FCFS) policy. Building on this, we model the operation of each witness as an M/H2/1 queue, which means that transactions arrive at the witness $w$ at a rate $\lambda_w$ and the service time is represented by a two-phase hyper-exponential distribution. With probability $p$, the first transaction in the queue receives service at rate $\mu_1$, while with probability $1-p$, it receives service rate at rate $\mu_2$. That is, the type of transaction is defined at the beginning of service. 
 
 The state of each witness is represented by a pair $(m,n)$, in which $m$ is the total number of transactions in the witness and $n\in\{1,2\}$ is the current service phase, which depends on the type of transaction being served. The stationary distribution of this queue in the witness $w$ can be obtained by Neuts' Matrix Geometric Method \cite{neuts}. We denote the stationary probability vector as:
 \begin{equation}
      \bm{\tau}^{(w)} = \left[\tau_0^{(w)}, \tau_1^{(w)}, \tau_2^{(w)}, \dotsc,\tau_k^{(w)}, \dotsc\right],
 \end{equation}
where $\tau_m^{(w)}$ is the steady-state probability of $m$ transactions in the witness $w$. Alternatively, the mean service rate is 
\begin{equation}
    \mu=\left(\frac{p}{\mu_1}+\frac{1-p}{\mu_2}\right)^{-1},
    \label{eq:mu}
\end{equation}
and the offered load to $w$ is $\rho_w=\lambda_w/\mu$. From there, the we calculate the variance of the service time
\begin{equation}
    \sigma^2_w=2\left(\frac{p}{\mu_1^2}+\frac{1-p}{\mu_2^2}\right)-\frac{1}{\mu^2}
\end{equation}
and the coefficient of variation $C_w^2=\mu^2\sigma_w^2$. Then, the average number of transactions in the queue of $w$ is
\begin{equation}
    L(w) = \sum_{m=0}^{\infty}m\tau_m^{(j)}=\rho_w+\left(\frac{1+C_w^2}{2}\right)\frac{\rho_w^2}{1-\rho_w}.
\end{equation}
Then, the number of \textit{local transactions} and \textit{Global transactions} handled by $w$ are, respectively,
\begin{equation}
    L_g(w) = pL(w)
\end{equation}
and
\begin{equation}
    L_l(w) = (1-p)L(w) = L(w) - L_g(w).
\end{equation}

\subsection{Global Blockchain (GB) System}
We model the GB as a modified $M/G^B/1$ queue as in~\cite{mgb1bc}. 
Let $L_g$ and $T_g$ be the RVs that define the number of 
transactions in the Blockchain queue and the confirmation time. We are interested in finding their mean values. For this, we define $b$ to be the maximum number of transactions in a block (i.e., the maximum block size). Hence, transactions are grouped into blocks and a new block is created when there are $b$ transactions in the Blockchain server.

Given that $p$ is the probability that a transaction sent to a witness is processed at the GB, the transaction arrival rate at the GB from the $v$ witnesses in the IoT deployment is 
\begin{equation}
    \lambda_B = \sum_{w=1}^{v} \lambda_w p.
    \label{eq:lambdaB}
\end{equation} 

We denote $U$ as  of the block generation time (i.e., the time it takes to generate a block) at the GB. Then, 

We define $U$ to be the continuous RV of the processing (i.e., service) time of a block at the GB. Hence, the system is stable and a limiting probability exists if and only if $\lambda_B E[U] <b$.

The cumulative distribution function (CDF) and the probability density function (pdf) of $U$ are denoted $G(x)$ and $g(x)$, respectively. We use these to calculate the hazard rate of $U$ as
\begin{equation}
    \theta(x) = \frac{g(x)}{1-G(x)}.
\end{equation}

Next, we define $L_g^s(t)$ as the number of transactions in server at time $t$, $L_g^q(t)$ as the number of transaction in the queue at time $t$, and $X(t)$ as the elapsed service time of the current transaction at $t$. From~\cite{mgb1}, we define 
\begin{multline}
    P_{m,n}(x,t)dx = \Pr\left[L_g^s(t)=m, L_g^q(t)=n ,\right.\\
    \left.x < X(t) \leq x+dx\right]
\end{multline}
to be the joint probability that, at time $t\geq 0$, there are $m\in\{0,1,2,...,b\}$ and $n\in\{0,1,2,...,x\}$ transactions in server and queue, respectively, and the elapsed service time lies between $x$ and $x+dx$. Next, we denote $P_{m,n}(x) = \lim_{t \xrightarrow{} \infty} P_{m,n}(x,t) $ and consider the two following cases. In the first one we have $\frac{d}{dx} P_{m,n}(x) = -\left[\lambda_B + \theta(x) \right] P_{m,n}(x) + \lambda_BP_{m,n-1}(x),
\text{ for } 0 \leq m \leq b \text{ and } n \geq 1,$ which shows that the number of transactions in the server and the queue does not change during a small interval. In the second one we have $\frac{d}{dx}P_{m,0}(x) = -\left[\lambda_B + \theta(x)\right]P_{m,0}(x), \quad \text{for } 0 \leq m \leq b,$ which occurs when a transaction arrives at the system with $0$ transactions in the queue. For the purposes of our study, it is sufficient to calculate the mean confirmation time as
\begin{IEEEeqnarray*}{rCl}
    E[T_g]\! &=&\! \Bigg[\lambda_B^2 E[U^2]-b(b-1)-2\left(b-\lambda_B E[U^2]\right) \\ 
    & &+ \sum_{n=0}^{b-1}\alpha_n\Big(\lambda_B E[U^2](b-n) +2bE[U](b-n)\IEEEyesnumber \IEEEeqnarraynumspace\\
    &&+E[U]\left(b^2-b-n^2+n\right)\!\Big)\!\Bigg] \frac{1}{2\lambda_B(b-\lambda_B E[U])},
\end{IEEEeqnarray*}
where $ \alpha_n = \sum_{m=0}^{b} \int \limits_{0}^{\infty} P_{m,n}(x) \theta (x) dx$. The interested reader is referred to~\cite{mgb1bc} for the fully detailed Blockchain queuing model.

\section{Performance Evaluation}
\label{sec:performance}

In this section, we use the queuing models described in Section~\ref{sec:analysis} to evaluate the performance of \emph{wiBlock} in terms of scalability. For this, we obtain the maximum transaction generation rate, along with the mean confirmation time and ledger size for both, the local and global Blockchain. We use the performance of a naive Blockchain and IoT integrated architecture, where the IoT devices communicate directly to the GB, as a benchmark. The mean results regarding the connectivity of the IoT devices with the witness system are obtained by a large number of Monte Carlo simulations and then used as an input to the queuing models. 

In our analysis, each device generates transactions at a rate $\lambda(i)= \lambda$ for all $i\in\mathcal{D}$. The block generation time $U$ is exponentially distributed with parameter $\mu_B= 1.8\cdot10^{-3}$~blocks per second. So we have $g(x)=\mu_B \exp(-\mu_B x)$, $E\left[U\right]=1/\mu_B$, and $E\left[U^2\right]=1/\mu_B^2$. Furthermore, we define the default block size to be $b=1000$~transactions. The rest of relevant parameters are listed in Table~\ref{tab:param}; these values are used unless otherwise stated.

\begin{figure}[t]
    \centering
    \includegraphics{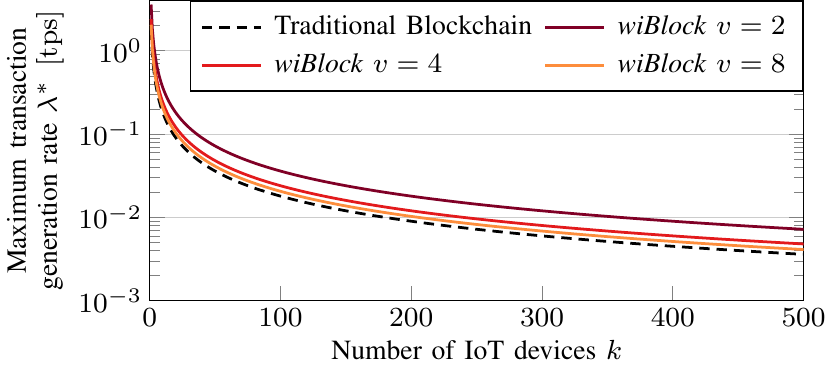}
    \caption{Maximum transaction generation rate per IoT device $\lambda^*$ for traditional Blockchain IoT and \textit{wiBlock} with random witness selection.}
    \label{fig:lambda_per_device}
\end{figure}

For the selected parameter settings, the GB system is stable when $\lambda_B^*=b/E\left[U\right]=1.8$. Building on this, from~\eqref{eq:lambdaB} we have that $\lambda<b/(k E\left[U\right])=1.8/k$ must hold for the GB to be stable in a traditional Blockchain architecture with $k$ identical IoT devices. Conversely, for \emph{wiBlock} with random witness selection, we have that only a fraction $p=(v-1)/v$ of the transactions must be sent to the GB. Hence, assuming no wireless channel errors occur and all the generated transactions are sent to a witness (i.e.,  $\sum_{w=1}^{v}\lambda_w=k\lambda$), the maximum load per IoT device that \emph{wiBlock} can handle is 
\begin{equation}
    \lambda<\frac{bv}{k E\left[U\right](v-1)}=\frac{1.8v}{k(v-1)}=\lambda^*(v),
\end{equation}
as shown in Fig.~\ref{fig:lambda_per_device} for $v=\{2,4,8\}$. 

\begin{table}[t]
\centering
\renewcommand{\arraystretch}{1.2}
\caption{Parameter settings for the performance evaluation.}
\begin{tabular}{ |lcl| } 
 \hline
 \textbf{Parameter} & \textbf{Symbol} & \textbf{Value} \\
 \hline
 \hline
Area of deployment & $A$ &  $100\times100~\mathrm{m}^2$ \\
\hline
Number of IoT devices & $k$ &  $500$  \\
\hline
Number of witnesses & $v$ &  $\{2,3,...,10\}$  \\
\hline
Carrier frequency&  $f$ & $914$ MHz \\
 \hline
Transmission power&  $P_t$ & $0.28183815$ W \\
 \hline
Antenna gains&  $G_t, G_r$ & $1$ \\
 \hline
Receiver sensitivity&  $\gamma$ & $3.652\cdot 10^{-10}$ W \\
\hline
Standard deviation of shadow fading & $\sigma_\text{dB}$ & $6$ dB \\[-0.2em]
 \hline
 Path loss exponent & $\beta$ & 3 \\\hline
Block size & $b$ & $1000$ transactions \\
 \hline
\end{tabular}
\label{tab:param}
\end{table}

\begin{figure}[t]
    \centering
    \includegraphics{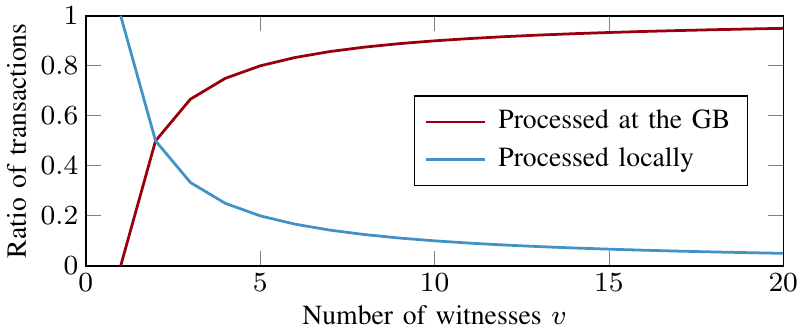}
    \caption{Ratio of transactions processed at the GB and at the witness system as a function of the number of witnesses $v$.}
    \label{fig:ratio_of_transactions}
\end{figure}

Hence, from the GB perspective, \emph{wiBlock} allows to deploy $1/p=v/(v-1)$ times more IoT devices than the naively integrated approach, as illustrated by Fig.~\ref{fig:ratio_of_transactions}. Note that the greatest gains in the scalability are obtained when $v$ is small, however, other factors such as the area coverage and processing capacity of the witness system must be taken onto account to select adequate values of $v$.

\begin{figure}[t]
  \centering
  \subfloat[]{
   \includegraphics[width=0.47\columnwidth]{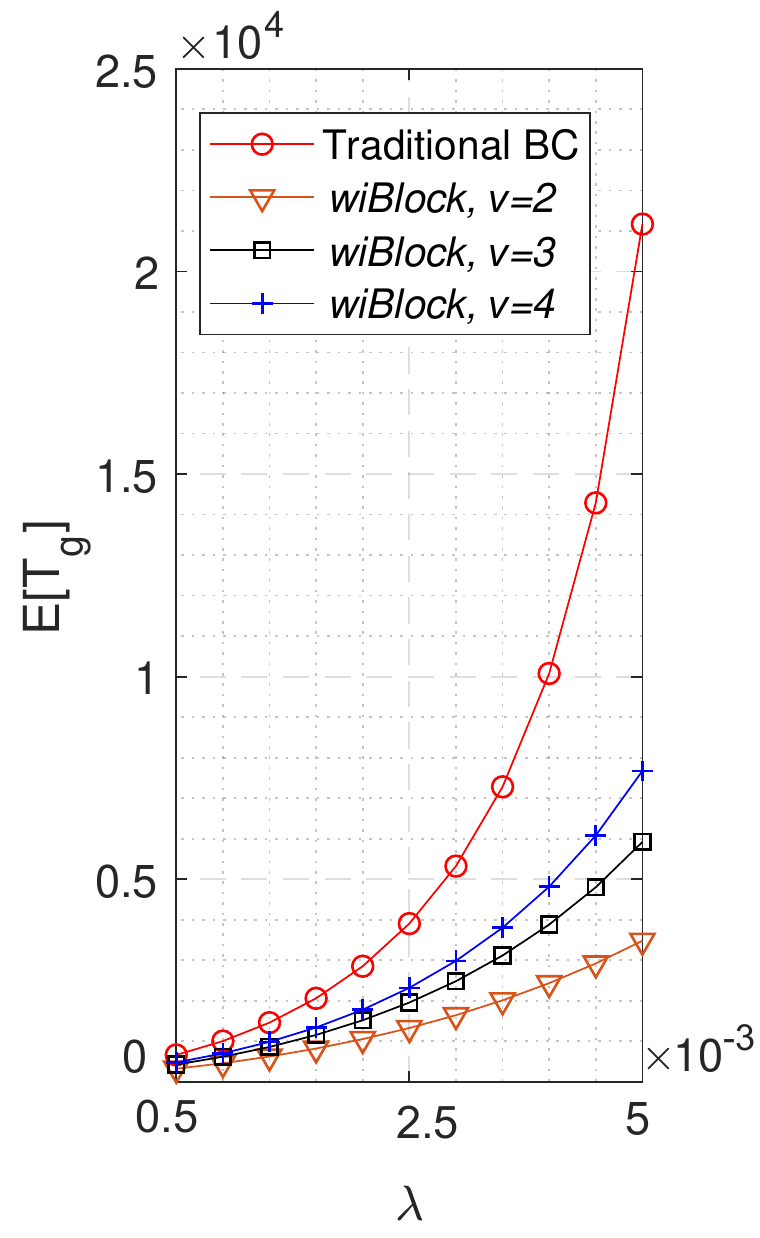}
    \label{fig:fig7a}}\hfil
    \subfloat[]{
    \centering\includegraphics[width=0.47\columnwidth]{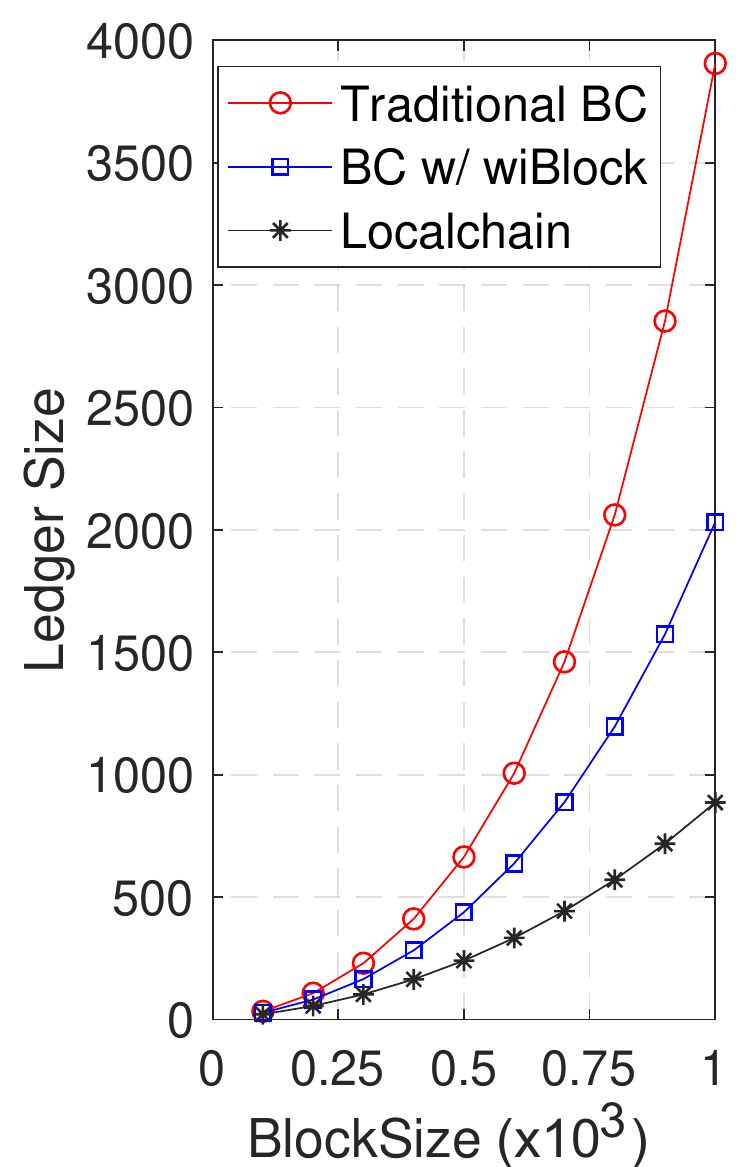}
    \label{fig:fig7b}}
  \caption{(a) Mean transaction confirmation time $E\left[T_g\right]$ as a function of the transaction generation rate at the IoT devices $\lambda$ and (b) ledger size at the GB and at each witness for $v=2$ as a function of the block size $b$ for traditional Blockchain IoT and \textit{wiBlock}.}
\end{figure}

Next, we evaluate the mean transaction confirmation time at the GB $E\left[T_g\right]$. Note that, in case a single witness is deployed in the system, all the transactions generated by the IoT devices will be considered as local transactions and processed locally. This can overload the witness, depending on its capabilities. In particular, the witness is stable if and only if the load offered to the witness is $\lambda_1<\mu_2$. Furthermore, deploying a single witness does not provide the necessary wireless coverage. That is, the more witnesses are deployed, the higher the probability of being able to communicate to, at least, one of them. Hence, we consider the cases where at least two witnesses are deployed, as shown in Fig.~\ref{fig:fig7a} for $v=\{2,3,4\}$.

As Fig.~\ref{fig:fig7a} shows, the witness system reduces the number of transactions sent to the GB and, as a consequence, greatly reduces the transaction confirmation time. Besides, the ledger size is considerably reduced, depending on the number of witnesses. This can be seen in Fig.~\ref{fig:fig7b} for $v=2$, where the ledger size of the GB is half of that with the traditional Blockchain and IoT integration, and the local ledger size at each witness is $1/v^2=1/4$ of it.

\section{Conclusion}
\label{sec:conclusions}
In this paper, we presented and evaluated the performance of a novel witness-based Blockchain system for IoT applications. As a starting point, we described the benefits of integrating Blockchain into IoT and the main challenges that must be overcome to achieve this integration. Building on these, we designed \emph{wiBlock}, an IoT-friendly distributed system that incorporates a witness system to address scalability issues of Blockchain. The scalability gains provided by \emph{wiBlock} are achieved by processing some of the transactions generated by the IoT devices locally, at the witness system. Our results show that the witness system greatly reduces the number of transactions transmitted to the Blockchain network and the transaction confirmation time. Future work includes the design of witness selection algorithms and implement \textit{wiBlock} in real testbed to further exploit the benefits provided by the witness system.

\section{Acknowledgement}
This  work  has  been  in  part  supported  by  the  European Research Council (Horizon 2020 ERC Consolidator Grant Nr.648382 WILLOW).

\bibliographystyle{IEEEtran}
\bibliography{IEEEabrv,bibliography}

\end{document}

%% file: figures/witness_selection.tikz
\begin{tikzpicture}[->, >=stealth', auto, semithick, node distance=1.4cm]
\tikzstyle{every state}=[draw=black,thick,text=black,scale=1,align=center, minimum size=13pt, inner sep=0, font =\footnotesize]
\tikzset{witness/.style={minimum size=13pt,rectangle, node distance=1.4cm}}

\coordinate (W) at (0,0);
\node[state, witness]    (w1) at (W) {1};
\node[state, witness]    (w2) [right of=w1] {2};
\node[state, witness]    (w3) [right of=w2] {3};
\node[node distance=1cm] (dw) [right of=w3]{$\cdots$};
\node[state,witness,node distance=1cm]    (wv) [right of=dw] {v};
\node[font=\footnotesize,anchor=west] at (-3,0) {Witness system $\mathcal{W}$};

\coordinate (D) at (0,1.3);
\node[state]    (d1) at (D) {1};
\node[state]    (d2) [right of=d1] {2};
\node[state]    (d3) [right of=d2] {3};
\node[node distance=1cm] (dd) [right of=d3]{$\cdots$};
\node[state,node distance=1cm]    (dk) [right of=dd] {k};

\node[font=\footnotesize,anchor=west] at (-3,1.3) {Set of IoT devices $\mathcal{D}$};

\path[thick,->]
(d1) edge (w1)
(d1) edge (w2)
(d1) edge (w3)
(d1) edge (wv);

\path[thick,->, OrRd-K]
(d2) edge (w1)
(d2) edge (w2)
(d2) edge (w3)
(d2) edge (wv);

\path[thick,->, OrRd-H]
(d3) edge (w1)
(d3) edge (w2)
(d3) edge (w3)
(d3) edge (wv);

\path[thick,->, OrRd-F]
(dk) edge (w1)
(dk) edge (w2)
(dk) edge (w3)
(dk) edge (wv);
\end{tikzpicture}